# MODELLING OF THE INFLUENCE OF NANOSTRUCTURES' SIZES ON LATTICE PARAMETERS

**M.A.Korshunov**

*L.V. Kirensky Institute of Physics Siberian Branch of Russian Academy of Sciences, Krasnoyarsk, Akademgorodok, 660036, Russia*
e-mail:mkor@iph.krasn.ru

Computer modeling of formation of the one-dimensional and three-dimensional monatomic nanostructures by the method of atom-atom potentials was done. The arrangement of atoms was defined on the basis of the energy minimum. Our calculations have shown that the distance between the nearest atoms depends on number of atoms, thus atoms on boundary of grains of nanoparticles have more loose packing than in the volume. In three-dimensional nanostructures at reduction of their sizes increase of the disorder of atoms on positions is observed. Examination of model nanostructures constructed of bi-atomic molecules was also carried out. It is shown that distance between the nearest molecules and their orientation depend on number of molecules surrounding them. Molecules on boundary of grains of nanoparticles have more friable packing than in the volume. As in the linear chain, and three-dimensional nanostructures constructed of bi-atomic molecules, not only the disorder on positions of molecules depending on the sizes of nanoparticles, but also the orientation disorder is observed. At increase of the sizes of nanostructures the lattice in volume becomes more ordered than on boundaries of nanoparticles grains.

In paper [1] the change of parametres of a lattice in nanocrystals attributed to the presence of vacancies in structure of grain boundaries. The stress on the grain boundaries leads to change of parameters of a lattice. In other paper [2], Sui et al. suggested that at reduction of the sizes of a nanocrystalline material the solubility of vacancies is considerably increases, which gives rise to a lattice expansion. If this is really so, crystalline lattice distortion should be uniformly distributed through the whole nanocrystal. However, the investigations show that crystalline lattice expansion exists mainly around the boundaries of the grain [3].

There are still no answer to a question: if the nanocrystal does not contain vacancies, whether change of parameters of a lattice will be observed?

To find an answer to this question we have made a calculations of lattice parameters of model of a nanocrystal depending on quantity of atoms. Calculations were done for both the linear chain and a three-dimensional case.

Interaction between atoms was described within the method of atom-atom potentials. At first, two atoms were considered and the energy minimization on distance between them was performed. After that, the third atom was added, and energy minimization on distances between three atoms was done. In a similar manner, all other atoms were added. In a three-dimensional case, this has defined structure of a lattice dependent on potential.

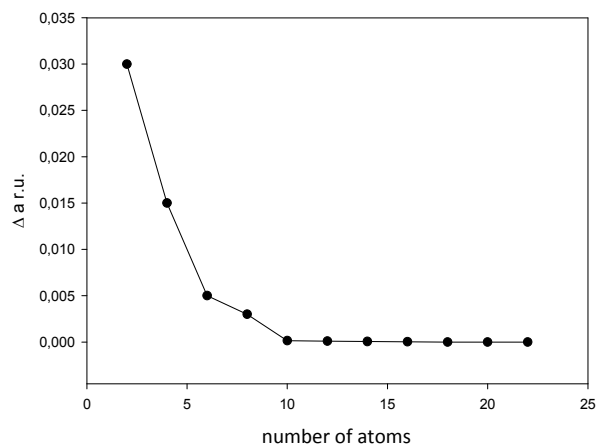

**Fig. 1.** Distance change between the nearest atoms Δa at centre of the linear chain depending on number of atoms.

For the linear chain it is found that at addition of atoms gradually there is a reduction of a lattice constant around the initial atom and a enlargement of the distance between atoms at the ends of the linear chain. In Fig. 1 the distance change Δa between the nearest atoms at centre of the linear chain with increase of number of atoms surrounding them is shown. The curve shape depends on the potential shape.

As one can see, at first, there is a sharp squeezing of a lattice and at the further addition of atoms squeezing becomes less pronounced. On edges of a chain squeezing is less than at centre (Δa~0.01r.u.), that is caused by interaction of border atoms only towards the chain centre.

For a three-dimensional case, similar effects are found, but the disordering of atoms on positions depending on number of neighboring atoms is observed. Less the number of atoms, the more disordering occurs. It is shown in Fig. 2. A deviation from an equilibrium position depends on a number of atoms (averaged on atoms the disorder in an arrangement of atoms in a plane of perpendicular to the considered interaction is shown). Besides, the disorder around grains boundaries is more than in the central region. From our calculations it is follows, that less the size of grain, the large the lattice expansion. And the lattice expansion is larger around the grain boundaries.

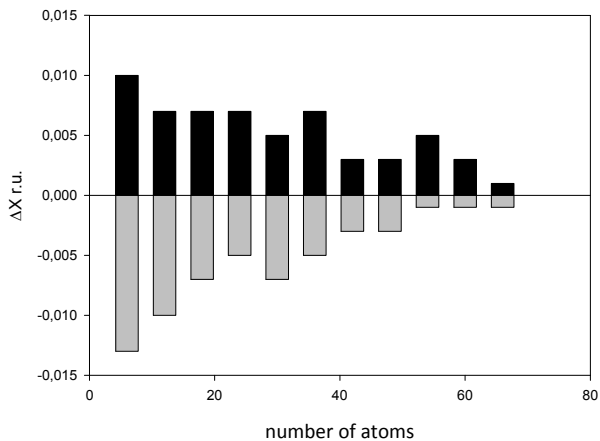

**Fig. 2.** A deviation from an equilibrium position of atoms ΔX in a three-dimensional case depending on number of atoms.

Hence, for atomic structures, the lattice expansion at reduction of the sizes of nanocrystals can take place without vacancies. The more the size of a nanoparticle the more lattice squeezing in the central region. However, around the grain boundaries smaller squeezing is observed. The disorder in atoms positions around equilibrium position also increase with reduction of the sizes of nanocrystals. At increase of the sizes of a nanocrystal the disorder in the central field decreases in comparison with grain boundary.

It is interesting to analyze whether the similar changes will be observed in case of nanostructures constructed of molecules.

We have made a calculations of parametres of a lattice of model of a nanocrystal consisting of bi-atomic molecules depending on their quantity. Calculations were done for both the linear chain and a three-dimensional case.

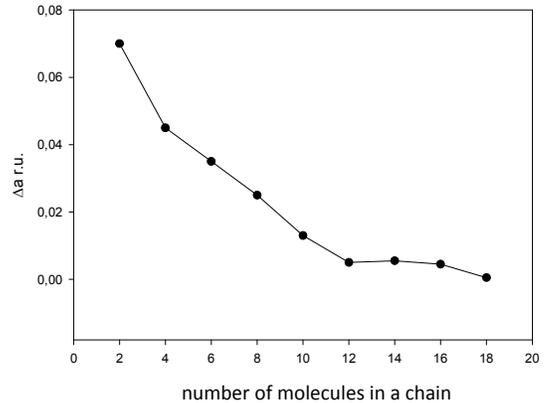

**Fig. 3.** Distance change Δa between the nearest molecules at centre of the linear chain with increase of number of molecules surrounding them.

Interaction between molecules was described within the method of atom-atom potentials. As the first step, we took two molecules and energy minimization on distance and mutual orientation between them was performed. In the next step, the third molecule was added, and energy minimization on orientations and distances between three molecules was done. Then, in the similar manner, the other molecules were added. In a three-dimensional case this defined the lattice structure.

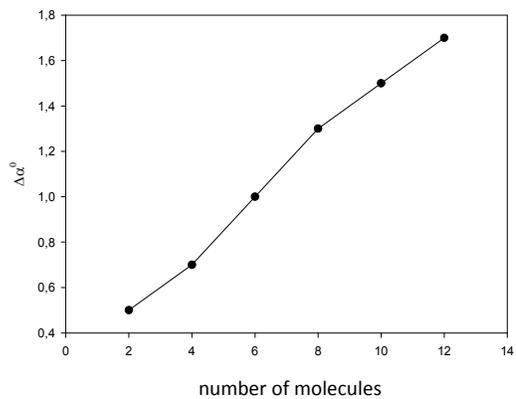

**Fig. 4.** Change of orientation of a molecule in chain centre $\Delta\alpha^0$ depending on number of molecules in it.

The calculations for the linear chain revealed that at addition of molecules gradually there is a reduction of a lattice parameter at centre of a chain and more friable packing at the boundaries of the linear chain. Orientation of molecules from centre to chain edges gradually changes. In Fig. 3 distance change Δa between the nearest molecules at centre of the linear chain with increase of number of molecules surrounding them is shown. Note, there is a lattice squeezing at addition of molecules, and on edges of a chain squeezing is less than at centre.

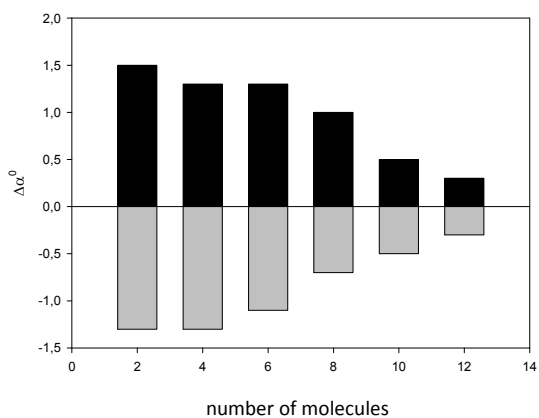

**Fig. 5.** Change of orientation of molecules in nanocrystal centre.

In Fig. 4 we show change of orientation of a molecule in chain centre $\Delta\alpha^0$ depending on number of molecules in it.

For a three-dimensional case, the similar effects were found. The disordering on positions and orientations of molecules depending on number of surrounding molecules is observed. Moreover, in a three-dimensional case the disorder of molecules depends on a direction.

In Fig. 5 the change of orientation of molecules at centre of a nanocrystal for one of directions is shown at change of number of molecules in a nanocrystal. Similar effects are found for the positions of centre of gravity of molecules.

Thus, it is shown, that more the size of a nanoparticle, the larger the lattice squeezing in the central region in comparison with grain boundaries. The disorder around the equilibrium position of molecules is also increases with reduction of the sizes of nanocrystals. The disordering on equilibrium positions and orientations of molecules depending on number of surrounding molecules is observed. Besides, in a three-dimensional case the disorder of molecules depends on a direction. The positional and orientation disorders at nanoparticle boundary to a less extent decreases with increase of quantity of molecules in a nanoparticle in comparison with the central part of a nanocrystal.